\documentclass[twocolumn,aps,amsmath,amssymb,showpacs]{revtex4}
\usepackage{graphicx}
\usepackage{epsfig}

\begin{document} 

\title{Anomalous synchronization threshold in coupled logistic maps}                
 
\author{C. Anteneodo}
\affiliation{Departamento de F\'{\i}sica, Pontif\'{\i}cia 
            Universidade Cat\'olica do Rio de Janeiro, 
            Caixa Postal 38071, 22452-970 RJ, Rio de Janeiro, Brazil}
\author{A. M. Batista}
\affiliation{Departamento de Matem\'atica e Estat\'{\i}stica, 
Universidade Estadual de Ponta Grossa, 84030-900 PR, Ponta Grossa, Brazil}
\author{R. L. Viana}
\affiliation{Departamento de F\'{\i}sica, 
            Universidade Federal do Paran\'a,  
            81531-990 PR, Curitiba, Brazil}

\begin{abstract} 
We consider regular lattices of coupled chaotic maps.   
Depending on lattice size, there may exist 
a window in parameter space  
where complete synchronization is eventually attained after a transient regime. 
Close outside this window, an intermittent transition to synchronization occurs. 
While asymptotic transversal Lyapunov exponents allow to determine 
the synchronization threshold,  
the distribution of finite-time Lyapunov exponents, 
in the vicinity of the critical frontier, is expected to provide 
relevant information on phenomena such as intermittency. 
In this work we scrutinize the distribution of finite-time exponents 
when the local dynamics is ruled by the logistic map $x \mapsto 4x(1-x)$. 
We obtain a theoretical estimate for the distribution of finite-time exponents, 
that is markedly non-Gaussian. The existence of correlations, 
that spoil the central limit approximation, is shown to modify 
the typical intermittent bursting behavior. 
The present scenario could apply to a wider class of systems with different local 
dynamics and coupling schemes. 
\end{abstract} 
 
 
 
\maketitle 


\section{Introduction}

Coupled map lattices (CMLs), dynamical systems with discrete 
space and time, are being intensively investigated since the early 80's as models 
of many spatiotemporal phenomena occurring in a wide variety 
of systems \cite{books}. 
One of such phenomena is synchronization and, in 
particular, amongst the various kinds of synchronized behavior, 
{\em complete synchronization} (CS) \cite{boccaletti} occurring 
in regular arrays of coupled chaotic systems. 
Intermittency plays an important role in the destabilization of the 
synchronized state\cite{intermit}. 
Intermittent transitions between laminar, quiescent states, and 
irregular bursting have been investigated in many dynamical systems, their 
scaling properties being determined\cite{int_teo,int_exp,ott}. 
Theoretical derivations of these scalings usually assume Gaussian fluctuations for 
describing the irregular bursts between consecutive laminar regions. 
Although this may be a suitable approximation for many systems, there are cases 
where deviations from Gaussianity can be observed, specially in the vicinity of 
the synchronization threshold. 
It is our purpose to exhibit such anomalous behavior.

Let us consider, as paradigmatic example of a spatially extended system, 
periodic chains of $N$ one-dimensional chaotic 
maps $x \mapsto f(x)$ evolving through a distance-dependent diffusive coupling: 
\begin{equation} 
x^{(i)}_{t+1}=(1-\varepsilon)f(x^{(i)}_{t})+\frac{\varepsilon} 
{\eta}\sum^{N'}_{r=1} 
B(r) \left( f(x^{(i-r)}_{t})+f(x^{(i+r)}_{t}) \right), 
\label{CML} 
\end{equation} 
where $x^{(i)}_t$ represents the state variable for the site 
$i$ $(i=1,2,...,N)$ at time $t$, $\varepsilon \ge 0$ is the 
coupling strength, $B(r)$ an arbitrary function of $r$ and 
$\eta=2 \sum^{N'}_{r=1}B(r)$ a normalization  
factor, with $N'=(N-1)/2$ for odd $N$.  

CS takes place when the dynamical variables that define the state of 
each map adopt the same value for all the coupled maps at each time 
step $t$, i.e., 
$x^{(1)}_t=x^{(2)}_t=\ldots=x^{(N)}_t\equiv x^{(*)}_t$.   
It can be easily verified that this state is solution of  
Eq. (\ref{CML}). The question is whether it is stable or not with respect to 
small perturbations in directions transversal to the CS state in CML phase space. 
A criterion for stability can be drawn from the asymptotic Lyapunov spectrum (LS). 
As far as the CS state lies along the direction associated to the largest 
exponent, the CS state will be transversely 
stable if the $(N-1)$ remaining exponents are negative.   
Therefore, if there is a single attractor, 
negativity of the second largest Lyapunov exponent   
implies that the CS state is asymptotically attained\cite{gade}. 
Depending thus on system size and on the parameters defining both the chaotic map and $B(r)$, 
there may exist an interval of values of the coupling strength $\varepsilon$  
for which the CS state is asymptotically stable. 
Explicit results have been shown before, for instance, 
for algebraically decaying interactions, i.e., 
$B(r)=1/r^\alpha$, with $0\le\alpha$\cite{apbv03,abv04} 
and for uniform interactions with a cut-off distance $\beta$, 
with $1\le\beta\le N'$\cite{gallas}. 
In those cases, for a given system size, 
one finds a {\em synchronization domain} in parameter space. 

Inside this domain, CS eventually occurs after a  transient whose typical duration  
diverges as one approaches the critical frontier.  
Outside the synchronization domain, there may occur intermittent 
behavior.    
A characterization of phenomena associated 
to a blowout bifurcation, such as intermittency\cite{ding_wang}, 
can be performed through the analysis of 
the distribution of largest transversal finite-time Lyapunov exponents (LTFEs). 
We will show that for Ulam maps coupled through schemes of the form 
(\ref{CML}), the probability density function (PDF) of 
LTFEs deviates from a Gaussian law. This will be shown to have important consequences 
at criticality.
Let us remark that analytical results are specially  relevant in this field due to the 
very nature of the phenomena involved, sensitive to the 
unavoidable finite precision of numerical simulations \cite{egs}.

\section{Finite-time Lyapunov spectrum}
\label{sec_ftls}

The dynamics of tangent vectors $\xi =(\delta x^{(1)},\delta x^{(2)},\ldots,$ 
$\delta x^{(N)})^T$ is obtained by differentiation of the 
evolution equations (\ref{CML}). In order to obtain finite-time exponents, 
we proceed analogously to what we have done before for the 
calculation of asymptotic exponents \cite{apbv03,abv04}.  
For self-containedness, let us review the steps. 
The tangent dynamics is given by $ \xi_{i+1}={\bf T}_i\xi_{i}$, 
where the Jacobian matrix ${\bf T}_i$ is   
\begin{equation} 
{\bf T}_i=\biggl( (1-\varepsilon)+\frac{\varepsilon} 
{\eta}{\bf B} \biggr){\bf D}_i, 
\label{tanmatrix} 
\end{equation} 
with the matrices ${\bf D}_i$ and ${\bf B}$ defined, 
respectively, by 
$D_{i}^{jk}=  f^\prime(x^{(j)}_i) \delta_{jk}$ and  
$B_{jk}= B(r_{jk}) (1-\delta_{jk})$ , 
being $r_{jk}=\mbox{min}_{l\in \cal{Z}}|j-k+lN|$. 
Once the initial conditions have been specified, the LS of finite-time 
Lyapunov exponents (FTEs) $\{ \bar\lambda_k(n) \}$, calculated over a time 
interval of length $n$, is extracted from the 
evolution of the initial tangent vector $\xi_0$:  
$\xi_n = {\bf \cal T}_n\xi_0$, 
where ${\bf \cal T}_n \equiv {\bf T}_{n-1}\dots 
{\bf T}_{1}{\bf T}_{0}$. 
The FTEs are obtained as $\bar\lambda_k(n)=\ln\bar\Lambda^{(k)}_n$, 
for $k=1,\ldots,N$, 
where $\{\bar\Lambda_n^{(k)}\}$ are the 
eigenvalues of $\hat{\Lambda}_n= ({\bf \cal T}^{T}_n  
{\bf \cal T}_n )^{\frac{1}{2n}}$\cite{ruelle}.

In CS states, the dynamical variables of all maps coincide at 
each time step $i$, i.e, 
$x^{(1)}_i=x^{(2)}_i=\ldots=x^{(N)}_i\equiv x^{(*)}_i$.   
In this case, ${\bf D}_i=f^\prime(x^{(*)}_i)\openone_{N}$, 
thus, ${\bf T}_i=f^\prime(x^{(*)}_i)\hat{\bf B}$  
and ${\bf \cal T}^{T}_i {\bf \cal T}_i = (\prod_{j=0}^{n-1} 
[f^\prime(x^{(*)}_j)]^{2}) \hat{\bf B}^{2i}$.  
Therefore, one arrives at 
the following expression for the spectrum of $N$ 
{\em finite-time} Lyapunov exponents, over a time interval of 
length $n$,  in {\em CS states}: 
\begin{equation} \label{lambdak}
\lambda_k(n)\,=\, \frac{1}{n}\sum^{n-1}_{i=0}\ln|f^\prime(x^{(*)}_i)| 
\,+\, c_k, \;\;\;\mbox{for $k=1,\ldots,N$},
\end{equation} 
with  $c_k \equiv \ln|1-\varepsilon +\varepsilon b_k/\eta|$, 
where $b_k$, the eigenvalues of the interaction matrix ${\bf B}$,  are
given by
$b_k=2\sum^{N^{\prime}}_{m=1}B(m)\cos(2\pi k m/N)$, for odd $N$. 
Here $x^{(*)}_i=f^i(x^{(*)}_0)$ is the $i$th iterate of the 
initial condition $x^{(*)}_0$, the same for the $N$ maps, since 
we are dealing with CS states. 

In the asymptotic case $n\to \infty$, and assuming ergodicity, 
the first term in the left-hand side of Eq. (\ref{lambdak}), 
which represents a time-average, can be substituted by an average 
over the single-map attractor.  In such case one gets the asymptotic LS  
\begin{equation} 
\lambda_k\,=\,\lambda_{U} \,+\, c_k,  
\label{asympt} 
\end{equation}
where $\lambda_{U}=\langle \ln|f^\prime(x^{(*)})| \rangle$ 
is the Lyapunov exponent of the uncoupled map. 
Notice that the parameters that define the particular  
uncoupled map affect only $\lambda_U$, while   
$\{ c_k \}$ are determined by the particular cyclic dependence on 
distance in the regular coupling scheme.

As discussed in the Introduction, 
the asymptotic LS provides a criterion for synchronization, 
namely the negativity of second largest (or largest transversal) 
asymptotic exponent $\lambda_\perp$.  
For the particular interaction $B(r)=1/r^\alpha$, that we will 
consider in numerical simulations, the above condition leads to\cite{apbv03,abv04}
\begin{equation} \label{critical} 
\frac{ 1-{\rm e}^{-\lambda_U} } 
 {1- b_1/\eta } < \varepsilon <
 \frac{ 1+{\rm e}^{-\lambda_U} }
 { 1- b_{N'}/\eta}.
\end{equation}

\section{Distribution of finite-time Lyapunov exponents}
\label{sec_distrib}

Unlike infinite-time exponents, local exponents strongly depend 
on the initial conditions. Starting from random initial conditions, 
the fluctuations in the values of $\lambda_k(n)$ arise from the 
summation in (\ref{lambdak}) only, since $\{c_k\}$ are constant. 
As a  consequence, all the $\lambda_k(n)$ 
of the  spectrum  will have a PDF with the same shape, 
differing only in the mean value  
$\langle \lambda_k(n) \rangle = \lambda_U+c_k$, which coincides  
with the asymptotic exponent $\lambda_k$. 
In particular the variance of finite-time Lyapunov exponents (FTEs) 
in the CS states does not depend on $c_k$ (since it is an additive constant). 
Therefore, it is the same for all the 
exponents of the spectrum, as expected also from the fact that the source 
of the fluctuations is unique. Along a trajectory, fluctuations shift 
the finite-time Lyapunov spectrum as a whole. 
A further consequence is that the variance of the local 
exponents in CS states does not depend on 
the lattice parameters embodied in $\{c_k\}$, but 
only on the local features of the individual map. 
In fact, in CS states, all maps evolve with the dynamics of an uncoupled map. 
Being so,  the PDFs of FTEs can be straightforwardly 
obtained from the PDF of the local map.
Hence, let us omit by the moment the index $k$. 
Our discussion will be valid for any $\lambda_k(n)$, in particular for
$\lambda_\perp(n)$.

For the Ulam map, $x\mapsto4x(1-x)$, a smooth approximate expression for the 
PDF of FTEs is available \cite{ulam}, namely
\begin{equation} \label{smooth}
P(\lambda(n)) \simeq \frac{2n}{\pi^2} 
\ln (\coth{| n[\lambda(n)-\lambda]/2|})\, , 
\end{equation}
valid for large enough $n$. 
For $\lambda(n)<\lambda$, expression (\ref{smooth}) is exact, however, for 
$\lambda(n)>\lambda$, the exact distribution presents a complex structure with 
$2^{n-1}$ spikes that get narrower and accumulate
close to the mean value with increasing $n$\cite{ulam,ramaswamy}. 
Therefore, in the latter interval,  
expression (\ref{smooth}) constitutes a smooth approximation, such  
that the sharp spikes have been trimmed by finite size bins. 
Moreover, the exact PDF 
is non-null only for $\lambda(n)-\lambda\le \ln2$. 

Even the smooth distribution (\ref{smooth}) is markedly different from Gaussian. 
It is divergent at $\lambda(n)=\lambda$ and falls off with 
exponential tails.  
It is noteworthy that the variance decays anomalously 
as $1/n^2$ \cite{ulam,ts95}
\begin{equation}
\sigma^2(n)=\frac{\pi^2}{6n^2}\left(1-\frac{1}{2^{n}}\right) \;,
\end{equation}
instead of the usual $1/n$ decay.  
Also notice that the PDFs (\ref{smooth}) for different values 
of $n$ would collapse into a single shape via rescaling by $n$.  
\begin{figure}[htb] 
\begin{center} 
\includegraphics*[bb=20 270 560 720, width=0.45\textwidth]{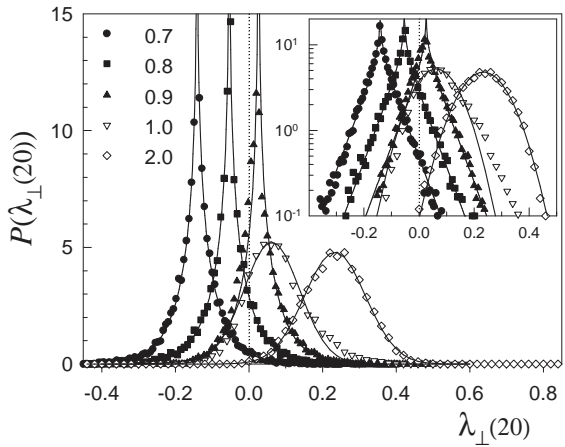} 
\end{center} 
\caption{Comparison between numerical and theoretical PDFs. 
They correspond to the largest transversal time-20  
exponent of $N=21$ coupled Ulam maps, with algebraically decaying interactions, 
for $\varepsilon=0.8$ and different values of $\alpha$ 
(above $\alpha_c\simeq 0.867$ stability of the CS state is lost). 
Numerical PDFs are represented by symbols. 
Solid lines associated to full symbols correspond to the theoretical prediction 
given by Eq. (\ref{smooth}), while those associated to hollow symbols correspond 
to Gaussian fittings. Inset: semi-log representation to exhibit the tails. 
} 
\label{fig_distr} 
\end{figure}

Fig.~\ref{fig_distr} exhibits numerical PDFs for CMLs together with the 
analytical prediction given by Eq. (\ref{smooth}). 
Numerical PDFs were built by choosing $10^4$ initial 
conditions and computing the second eigenvalue of the matrix $\hat{\Lambda}_n$ 
after a transient. 
Eq. (\ref{smooth}), obtained for uncoupled maps, is in excellent agreement with 
numerical results for LTFEs in the coherent states of CMLs, as expected. 
It is worth noting that, although expression (\ref{smooth}) was derived for CS states,  
it remains still a good approximation even close  
outside the synchronization domain.
This means that the correlations that lead to violation of the central limit theorem 
persist in that region. 
However, far enough from the threshold, the terms that 
contribute to FTEs become 
uncorrelated by the ``bath'' of coupled chaotic maps, 
and the central limit theorem holds, leading to Gaussian shapes 
(see Fig.~\ref{fig_distr}).  
Notice that the PDF for $\alpha=1.0$, although Gaussian in the central part, 
stills falls at right with a exponential tail. 
A detailed analysis of correlations giving rise to (\ref{smooth}) 
can be found in Refs. \cite{ulam,ts95}.
Basically, the autocorrelation function $C(n)$ of one-step FTEs is 
$C(n)=-\pi^2 2^{-n}/24$, that is, it decays exponentially with time $n$ (with a 
characteristic time equal to the inverse of $\lambda_U=\ln 2$). 
However, since successive one-step exponents are dependent through the 
deterministic logistic mapping, non-Gaussian PDFs of FTEs arise.

\section{Consequences of fluctuating exponents}
\label{sec_conseq}

\subsection{Subcritical regime}

\begin{figure}[htb] 
\begin{center} 
\includegraphics*[bb=80 320 540 650, width=0.45\textwidth]{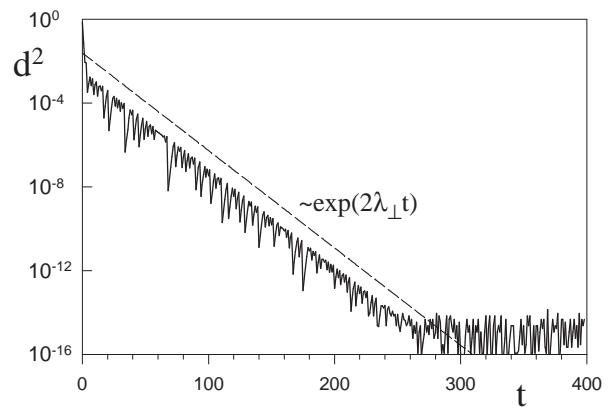} 
\end{center} 
\caption{Relaxation to the synchronization manifold. 
Time series of $d^2$ for the same CML of Fig.~\ref{fig_distr} with  $\alpha=0.8$ (subcritical).  
The dashed line corresponds to the exponential law indicated in the figure, for comparison.
} 
\label{fig_sub} 
\end{figure}

For parameter values belonging to the synchronization domain, 
that is, for subcritical $\alpha$, being all other parameters fixed, 
the system eventually converges to the CS state.
In fact,  asymptotically,  CS states are stable since the
distribution  of LTFEs collapses to a Dirac delta function centered at  
$\lambda_\perp$, which is negative in that domain. 
The relaxation to a CS state can be measured, for instance, 
by means of either the distance to 
the SM, defined through $d(t)=\sqrt{ \sum_i(x^{(i)}_t-\langle{x}_t\rangle)^2 }$, or the 
order parameter $R(t)=|\sum_i\exp(2\pi x^{(i)}_t)|/N$. 
Since, for small deviations from the SM, both quantities are related through 
$d^2\simeq (1-R^2)/2$, we will exhibit the time evolution of $d^2$ only. 
After a very brief transient, the decay to the CS state is exponential   
with a characteristic time  given by $\tau_c=1/|\lambda_\perp|$ 
(see Fig.~\ref{fig_sub}),  
that diverges at the critical frontier. 
For the power-law interaction, $\lambda_\perp$ scales as 
$|\lambda_\perp|\sim |\alpha-\alpha_c|$ 
and $|\lambda_\perp|\sim |\varepsilon-\varepsilon_c|$, 
at the critical point. 

The fact that the distribution of LFTEs spreads over negative and positive values 
(Fig.~\ref{fig_distr}), 
implies that the exponents, computed over finite-time segments of a trajectory, 
fluctuate around zero.  
On one hand, as one approaches the frontier subcritically, 
the mean value of the distribution shifts to zero from negative values. 
On the other hand, as one follows a trajectory for a longer time 
interval, the PDF of LTFEs concentrates around the mean. 
Then, there may be segments of trajectory in which the lattice is repelled from the 
synchronization manifold (SM). 
But, in average, it is attracted exponentially fast. Due to the finite precision of 
computer calculations, the distance to the SM saturates (see Fig.~\ref{fig_sub}). 
Intrinsic noise, due to numerical truncation, may drive the state of the system slightly away 
from the saturation level. However, 
each time this happens, the distance decays, again exponentially fast, 
to its lower bound.

Numerical results in the subcritical regime, illustrated in Fig.~\ref{fig_sub}, 
can be suitably described by $d(t)=d_o \exp[\lambda_\perp(t)\,t]$. Then, from 
Eq. (\ref{lambdak}), we have 
\begin{equation} \label{distance}
d(t)\;=\; d_o\exp[\lambda_\perp \,t + \sum_{n=0}^t \zeta(n) ]
\end{equation} 
where we have split the fluctuating component $\zeta=\lambda_\perp(1)-\lambda_\perp$, 
corresponding to successive (time-correlated) centered one-step LTFEs. 
The evolution of the distance to the SM can also be modeled by a It\^o 
multiplicative stochastic differential equation for $d$, 
\begin{equation}\label{ito}
\dot{d}=\lambda_\perp d+\zeta(t)d \,,
\end{equation}
where $\zeta(t)$ is a colored non-Gaussian noise and time is continuous.  
 
\begin{figure}[htb] 
\begin{center} 
\includegraphics*[bb=100 200 508 650, width=0.45\textwidth]{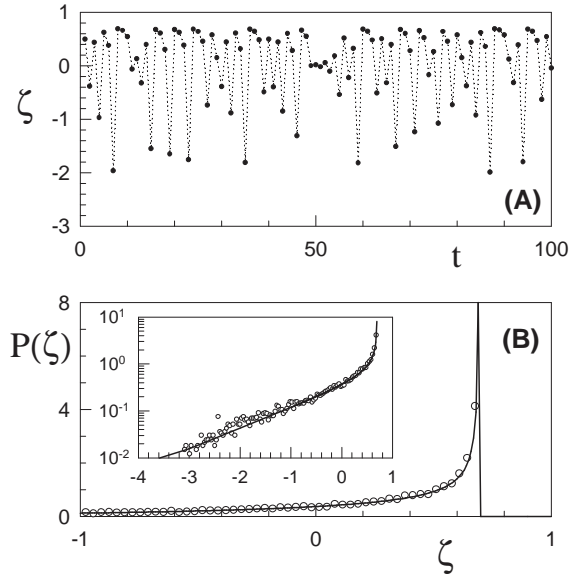} 
\end{center} 
\caption{Time evolution of $\zeta=\lambda_\perp(1)-\lambda_\perp$ (A)
for the same CML of Fig.~\ref{fig_distr} with  $\alpha=0.9$ (supercritical). 
Normalized histogram of $\zeta$ (B). Inset: semi-log representation to exhibit the 
exponential tail. Full lines correspond to Eq. (\ref{one}). 
} 
\label{fig_zeta} 
\end{figure}

As shown in Sect. \ref{sec_distrib}, in particular, 
the distribution of $\lambda_\perp(1)$ in synchronized states 
follows that of one-step FTEs in the uncoupled map. 
An exact expression can be straightforwardly 
derived from the invariant measure of the Ulam 
attractor\cite{ramaswamy}. Then, one has  
\begin{equation} \label{one}
P(\zeta) =\frac{2}{\pi}\frac{1}{\sqrt{4{\rm e}^{-2\zeta}-1}} \, , 
\end{equation}
with $\zeta\in(-\infty,\ln2)$.  
Notice that $\zeta$, with zero-mean, is bounded from above. This   
explains the upper bound of the fluctuations superposed to the exponential decay 
in Fig. (\ref{fig_sub}). In fact, the fluctuations around the exponential envelope  
can be fully described by the statistics of time-one exponents. 
The time series and distribution of $\zeta$ are presented in Fig.~\ref{fig_zeta} 
for supercritical $\alpha$, showing that even close outside the synchronization domain, 
the distribution of $\zeta$ follows that of the uncoupled map.

Breakdown of shadowability might occur\cite{shadow}, as 
exponents fluctuating about zero are a signature of unstable dimension variability.
Taking into account that the PDF is non-null for 
$\lambda_\perp(n) \le \lambda_\perp +\ln 2$\cite{ulam,ramaswamy}, then 
if $\lambda_\perp(\alpha,\varepsilon,N)<-\ln 2$, the 
finite-time exponents are negative for almost any initial condition.  
Thus, this point would correspond to the onset of shadowability. 
Whereas, for $-\ln 2<\lambda_\perp$, although the mean of the distribution may be 
negative, there is always a non-null fraction $f$ of positive exponents  given by 
$f= \int_0^\infty {\rm d} \lambda_\perp(n) P(\lambda_\perp(n))$,  
pointing to the possibility of loss of shadowability of numerical trajectories\cite{review}. 
Because $f$ grows from zero with a very small slope, since the positive tail 
of the PDF is approximately exponential, then the onset may appear shifted towards 
the threshold in numerical computations\cite{validity}.

\subsection{Supercritical regime}

\begin{figure}[htb] 
\begin{center} 
\includegraphics*[bb=80 320 540 700, width=0.45\textwidth]{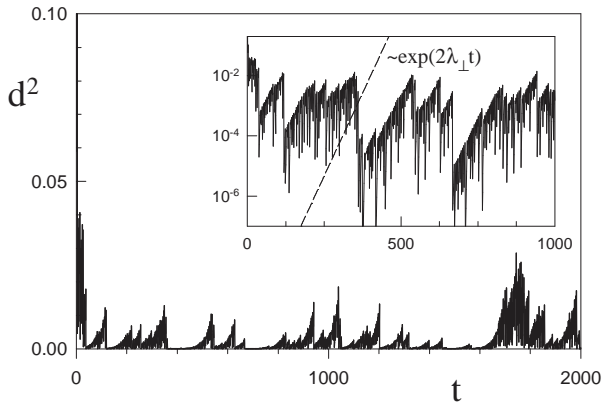} 
\end{center} 
\caption{Time series of $d^2$   
for the same CML of Fig.~\ref{fig_distr} with $\alpha\simeq 0.9$ (supercritical). 
Inset: semi-log representation of the same data. 
The dashed line corresponds to the exponential law indicated in the figure, for  
comparison.
} 
\label{fig_super} 
\end{figure}

For supercritical $\alpha$, that is outside the synchronization domain, 
CS states are not asymptotically stable, because $\lambda_\perp>0$. 
Close to the threshold (up to $\alpha\simeq 1$ for $\epsilon=0.8$), 
the numerical estimate for the second largest transversal exponent 
fairly coincides with $\lambda_\perp$ (calculated for synchronized states). 
Moreover, close enough to the boundary, 
correlated bursts away from the SM occur (see Fig.~\ref{fig_super}). 
Although this figure exhibits a time series up to $t=2000$, the same features 
are observed for longer runs (performed typically up to $t=10^7$).

The intermittent behavior can also be understood in terms of the distribution 
of the finite-time largest transversal exponent $\lambda_\perp(n)$, that close 
to the threshold can be described by Eq. (\ref{smooth}). 
Although the average $\langle \lambda_\perp(n)\rangle = \lambda_\perp$ 
is positive, there is a non-null probability that the LTFE  
be negative, thus leading to intermittent behavior at finite times. 
The distribution of LTFEs indicates that there are time intervals during 
which the trajectories are either attracted to or repelled from the SM.  
Despite the average duration of the synchronized time intervals increases when approaching 
the critical frontier, synchronization is not attained as a 
final stable state.

Intermittent bursts grow exponentially with a characteristic time given by $1/\lambda_\perp$ 
(see inset of Fig.~\ref{fig_super}). Therefore, the characteristic time has 
the same scaling laws as 
the subcritical characteristic time discussed above. 
The distance to the SM fluctuates around a reference level 
($\langle d\rangle\simeq 0.037 $ for the parameters in Fig.~\ref{fig_super})  that 
increases with $\alpha$.
Nonlinearities in Eq.~(\ref{ito}) cause saturation of the growth experienced by the 
distance to the SM due to the positiveness of $\lambda_\perp$, and keep the distances
within a bounded interval. 
Close to the threshold, the average distance increases   
linearly with $\lambda_\perp\sim|\alpha-\alpha_c|$. 
Also, for increasing $\alpha$, the correlated bursts become more frequent (hence, 
its duration becomes shorter) such that 
far from the frontier fluctuations become uncorrelated and the 
intermittent clustering effect disappears.  
Moreover, in that region, the numerical estimate of the largest transversal 
exponent significantly deviates from the one calculated for synchronized states 
and new features, out of the scope of the present work, occur. 
Therefore, we will focus on the near vicinity of the threshold.  

\begin{figure}[htb] 
\begin{center} 
\includegraphics*[bb=60 380 520 670, width=0.45\textwidth]{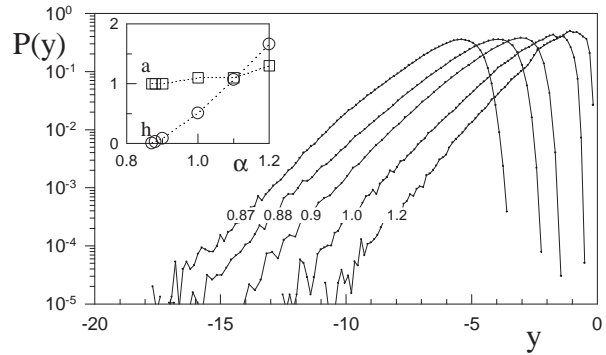} 
\end{center} 
\caption{Distribution of the logarithmic distance $y=\ln d$, for different values 
of $\alpha$. Inset: exponent $a$, resulting from the 
best exponential fit to the initial increasing regime, 
and hyperbolicity exponent $h$ and as a function of $\alpha$.   
} 
\label{fig_dista} 
\end{figure}

The histogram  of logarithmic distances $y\equiv\ln d$ is presented in Fig.~\ref{fig_dista}. 
The distribution initially grows approximately as $\exp(ay)$ and 
above the maximum value falls off faster than exponentially. 
The coefficient of the exponential argument is $a\simeq 1$, at variance, with the value 
$a=2\lambda_\perp/\sigma_1^2\equiv h$ (hyperbolicity exponent)\cite{ott}, 
where $\sigma^2$ is the variance of time-1 exponents, 
derived under assumption of Gaussian fluctuations. 
In fact, in the continuum approximation, the evolution of $y\equiv\ln d$ follows, 
at first order, the Langevin equation

\begin{equation} \label{langevin}
\dot{y}=\lambda_\perp + \zeta(t) \,
\end{equation}
where $\zeta$ is a fluctuating quantity with zero-mean and variance $\sigma_1^2$. 
If fluctuations were white Gaussian ones,  
then, from the stationary solution of its associated Fokker-Planck equation, 
the distribution of $y$ should increase following the law $P(y)\sim \exp(hy)$, 
with $h$ defined as above. 
Multiplicative corrections to Eq. (\ref{langevin}), 
to model the upper-bounded behavior of $d$, 
will affect the shape of $P(y)$ mainly above the average value, and are not expected to 
affect the small $d$ behavior. 
The main point is that the stochastic process $\{\zeta\}$ is not 
white nor Gaussian, as can be neatly observed in Fig. \ref{fig_zeta},  
what may explain the deviation from the $\exp(hy)$ law. 

\begin{figure}[htb] 
\begin{center} 
\includegraphics*[bb=60 400 550 705, width=0.45\textwidth]{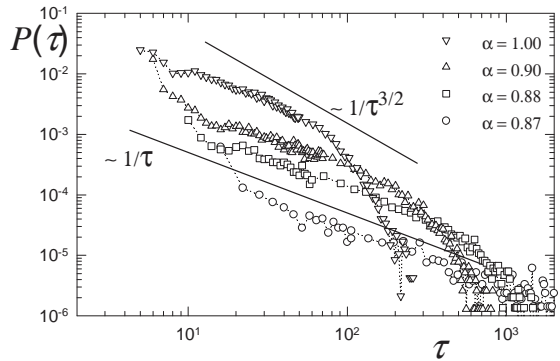} 
\end{center} 
\caption{Distribution of inter-burst times 
for the same CML of Fig.~\ref{fig_distr} with different (supercritical) values of 
$\alpha$ indicated on the graph. In all cases, $d_o\simeq 2\langle d\rangle$, but 
we verified that decay laws do not substantially 
depend on the choice of the threshold $d_o$. 
Full lines, corresponding to the indicated power-laws, were drawn for comparison. 
} 
\label{fig_bursts} 
\end{figure}

A universal result is that, at the onset of intermittency, the distribution of laminar phases 
(inter-burst intervals) decays as a power-law\cite{int_teo}, 
meaning the presence of lengths of arbitrarily large size. 
In particular the size of the average plateau diverges. 
Moving far from the onset, the tail of the distribution of laminar phases 
is gradually dominated by  an exponential decay.  
In general, the distribution of lamellar phases can be obtained by 
solving a first-return problem. 
For the power-law decay, the exponent $\beta=3/2$ was found to 
be universal, as far as the central limit approximation holds\cite{int_teo}. 
However, as we have seen, in our case, 
very close to the threshold, correlations persist and the distribution of finite-time 
exponents, even, and specially, in the central part (where it is divergent), 
deviates from the Gaussian approximation. Therefore, deviations from the 3/2 power 
law are expected. 
We measured inter-burst sizes, that is, the length of time segments during which 
the distance $d$ remains below a threshold value $d_o$. 
Numerical  distributions of inter-burst sizes, for 
values of parameter $\alpha$  close  to the synchronization threshold, 
are displayed in Fig.~\ref{fig_bursts}. 
In general, a rapid decay is observed for very small $\tau$ (not exhibited).
Since a power-law behavior is expected for sufficiently large $\tau$, 
histogram heights for small values of $\tau$ are not exhibited in the figure, 
for the sake of clearness.  
One sees that the PDF of inter-burst intervals follows a power-law 
with $\beta\simeq 1$, at neat variance with the value $\beta= 3/2$.  
A cross-over to a asymptotic exponential regime is always observed.
We have already seen in Fig.~\ref{fig_distr} that, below $\alpha\simeq 1$, 
correlations persist. This may explain 
the power-law exponent observed in Fig.~\ref{fig_dista}.

\section{Summary}

We have presented analytical results for the PDF of LTFEs in CS 
states of coupled Ulam maps. 
They were confirmed by numerical experiments performed for 
CMLs with interactions decaying with distance as a power law. 
The knowledge of the statistical properties of finite-time Lyapunov exponents 
allows to understand the anomalies in bursting behavior close to the SM. 
As a consequence, universal laws derived under assumption of Gaussian fluctuations 
do not hold generically. 

Our results were obtained for local Ulam maps, however, 
the observed features may occur in a wider class of extended systems, as 
long as other deterministic chaotic or stochastic processes present similar deviations 
from Gaussianity\cite{ramaswamy}.

\section*{Acknowledgments}
This work was partially supported by Brazilian agencies CNPq and
Funda\c{c}\~ao Arauc\'aria.

\end{document}